\documentclass[conference]{IEEEtran}
\IEEEoverridecommandlockouts
\usepackage[utf8]{inputenc}

\usepackage{url}
\usepackage{todonotes}

\newcommand{\eg}{e.\,g.,\ }
\newcommand{\ie}{i.\,e.,\ }

\usepackage{csquotes} 

\usepackage{amssymb, amsmath, MnSymbol}
\usepackage{amsthm}
\usepackage{microtype}
\usepackage{enumerate}

\usepackage{fontawesome}
\usepackage{svg}
\usepackage{hhline}

\usepackage{float}
\usepackage{listings}
\newfloat{lstfloat}{htbp}{lop}
\floatname{lstfloat}{Listing}

\newcommand{\qedblack}{\hfill \ensuremath{\blacksquare}}

\newcommand{\labeltitle}[1]{\vskip 0.03in \noindent\textbf{#1}} 

\theoremstyle{definition}
\newtheorem{example}{Example}

\usepackage[capitalise,nameinlink]{cleveref}
\crefname{section}{Sect.}{Sects.}
\Crefname{section}{Section}{Sections}
\crefname{figure}{Fig.}{Figs.}
\Crefname{figure}{Figure}{Figures}
\crefname{table}{Tab.}{Tabs.}
\Crefname{table}{Table}{Tables}
\crefname{example}{Ex.}{Exs.}
\Crefname{example}{Example}{Examples}

\definecolor{amber}{rgb}{1.0,0.75,0.0}

\begin{document}
\title{Towards Automated Attack Simulations of BPMN-based Processes}

\author{\IEEEauthorblockN{Simon Hacks, Robert Lagerström}
    \IEEEauthorblockA{Division of Network and Systems Engineering\\
    KTH Royal Institute of Technology\\
    Stockholm, Sweden\\
    \{shacks$|$robertl\}@kth.se} \and
    \IEEEauthorblockN{Daniel Ritter}
    \IEEEauthorblockA{
        SAP SE\\
        Walldorf (Baden), Germany\\
    	daniel.ritter@sap.com}
    }
\maketitle

\begin{abstract}
    
    
Process digitization and integration is an increasing need for enterprises, while cyber-attacks denote a growing threat.
Using the Business Process Management Notation (BPMN) is common to handle the digital and integration focus within and across organizations.
In other parts of the same companies, threat modeling and attack graphs are used for analyzing the security posture and resilience.

In this paper, we propose a novel approach to use attack graph simulations on processes represented in BPMN.
Our contributions are the identification of BPMN's attack surface, a mapping of BPMN elements to concepts in a Meta Attack Language (MAL)-based Domain-Specific Language (DSL), called coreLang, and a prototype to demonstrate our approach in a case study using a real-world invoice integration process.
The study shows that non-invasively enriching BPMN instances with cybersecurity analysis through attack graphs is possible without much human expert input.
The resulting insights into potential vulnerabilities could be beneficial for the process modelers.
\end{abstract}


\section{Introduction}
%
In a highly connected world, in which enterprises get more and more intertwined with each other, digitization drives the change of our society, which enables innovation, increased connectivity, improved productivity, and more accessible information \cite{jeske2020achievements}. 
Many organizations move their applications into the cloud, integrate them within the organization and across different organizations \cite{DBLP:journals/is/RitterMR17}.
To facilitate the related internal and inter-organizational business and application integration processes, many organizations use the Business Process Management Notation (BPMN) \cite{DBLP:conf/ecmdafa/Ritter14,DBLP:conf/caise/0001H15,DBLP:journals/ijcis/RitterS16}.
At the same time, we perceive a significant growth of cyber-attacks \cite{jangjaccard2014a,mosteanu2020challenges}.
A potential approach to address these security threats is an assessment of the integrated processes from a cyber security perspective.

Different ways to enrich the standardized BPMN with new concepts that reflect various security aspects already exist (\eg \cite{maines2015cyber,meland2012representing,mulle2011security,salnitri2014modeling,zareen2020security}).
However, if organizations have a large set of productive processes, adding supplementary information is often an unfeasible effort. Moreover, those processes are often created by business experts and process consultants, who regularly have no security expertise.
Even when assuming that process-aware runtime systems are routinely assessed regarding potential security threats, scenario-specific configurations, and design decisions (\eg selection of server or library versions) could make instantiated processes vulnerable to attacks.
Finding all possible vulnerabilities and avoiding exploits is extremely difficult for process modelers and project consultants.
At the same time, exploits could have a huge impact across overlapping or connected processes, and thus endanger larger parts of a business.

For example, the BPMN-based integration process in \cref{ex:invoicing} exhibits several points of attack that might not be obvious during its scenario-specific instantiation.
\begin{example}
\label{ex:invoicing}
%
\begin{figure*}[bt]
	\centering
	\includegraphics[width=0.8\linewidth]{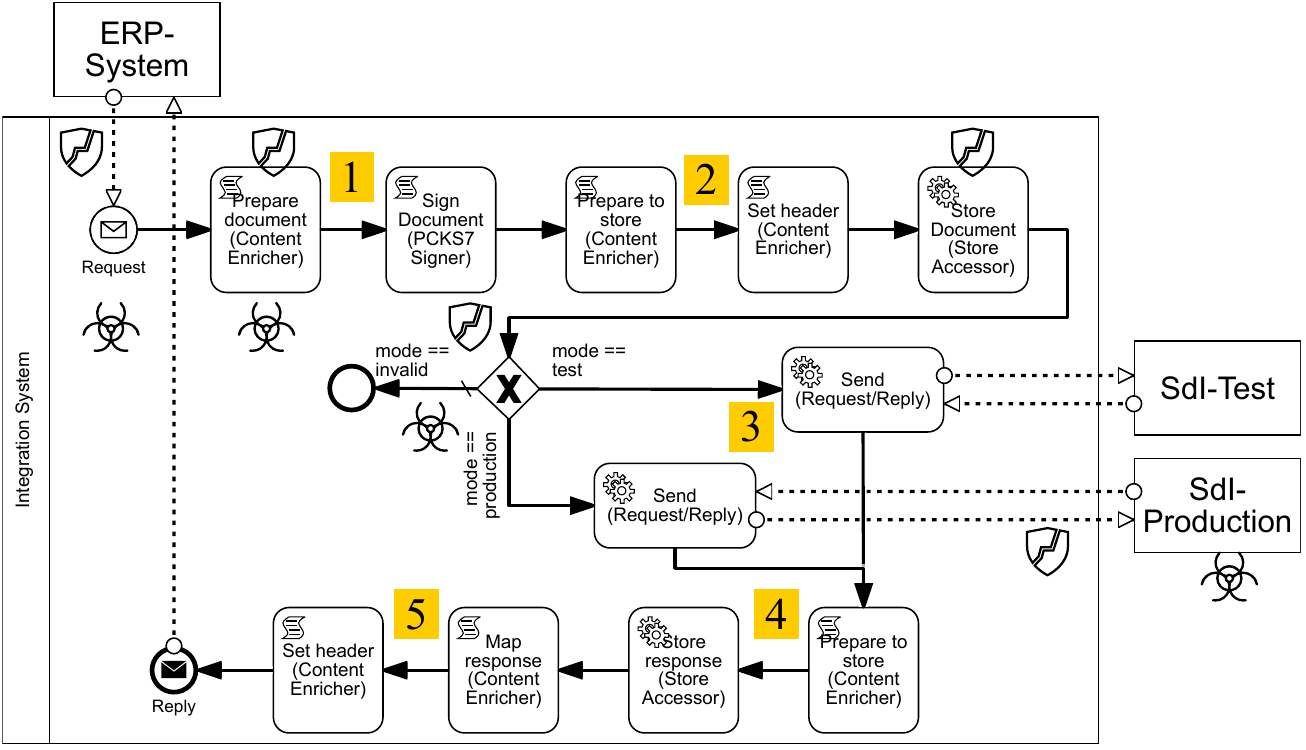}
	\caption{\label{fig:cdd}SAP CPI eDocuments Italy invoicing integration process (simplified from \cite{sap-hci-content}) with non-mandatory, explanatory vulnerabilities and exploits 
	}
\end{figure*}
\Cref{fig:cdd} shows a simplified version of an invoicing integration process for Italian companies.
An ERP system sends invoices to the integration process, which \textbf{{\colorbox{amber!70}1}} preserves information like the transfer \textsl{mode}, as well as relevant identifiers like \textsl{IdentityCode} (data exchange not shown), and then signs the invoice itself. 
In \textbf{{\colorbox{amber!70}2}} a Sistema di Interscambio (SdI)\footnote{SdI is an Italian platform for issuing, transmitting, and saving invoices.} compliant message is prepared 
and the signed invoice is cached, before storing it for further reference.
Message headers required by SdI are set 
and \textbf{{\colorbox{amber!70}3}} either sent to a test or production endpoint, else discarded, depending on its mode.
The response message is enriched with information preserved from the original message 
\textbf{{\colorbox{amber!70}4}}, stored for reference and then \textbf{{\colorbox{amber!70}5}} a compliant response message for the ERP system is prepared.
\qedblack
\end{example}
Starting with the data transfer from the ERP system to the integration process in the form of a message, the remote call is conducted over public networks to a scenario-specific endpoint configuration (\eg HTTPS server).
Such configurations are prone to security vulnerabilities \includegraphics[scale=0.15]{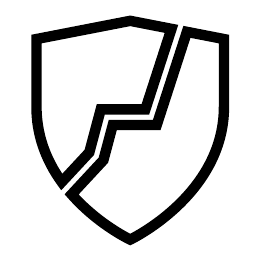}, potentially leading to an exploit \includegraphics[scale=0.023]{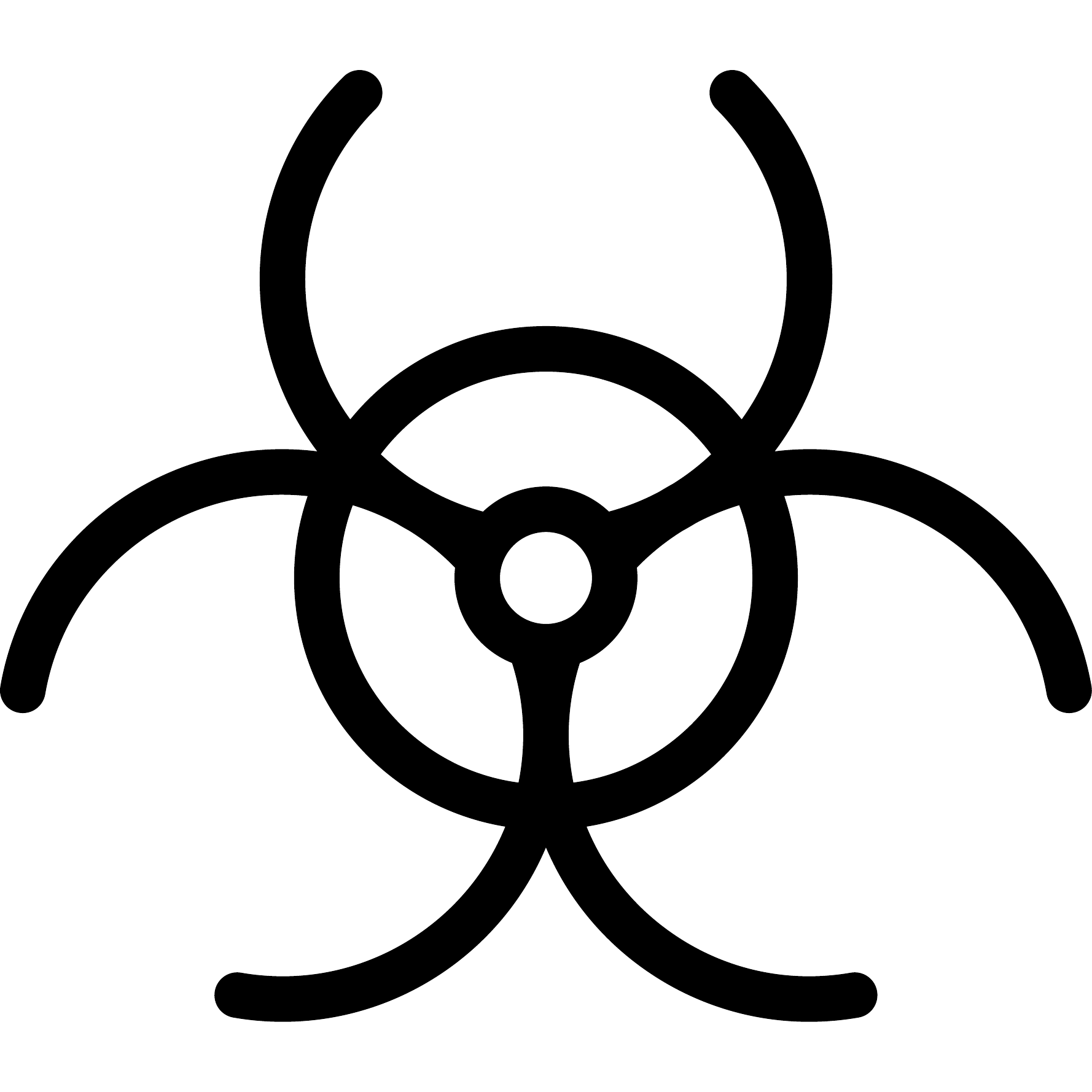} of the integration process runtime.
Similarly, subsequent script steps allow for user-defined source code that could exploit vulnerabilities of the script engines, and consequently the integration runtime.
Service tasks accessing an associated, global data store (not shown), as well as remote calls to the SdI endpoints could lead to attacks on the connected database, remote server endpoints, respectively.
Finally, the user-defined conditions in the exclusive gateway could exploit weaknesses in the used parser.

We aim to develop a solution that allows to assess the security state of BPMN processes without the need to edit the process models.
To perform such a security assessment, one has to solve the challenging tasks of identifying vulnerabilities, understanding security-relevant parts, and determining potential attacks \cite{6041993}.
The use of attack simulations based on system architecture models have been proposed for addressing these challenges, (\eg  \cite{ekstedt2015securi,HolmSBE15}). 
These approaches have in common that they rely on a static implementation of the used meta-model.
Therefore, the use of the Meta Attack Language (MAL) \cite{johnson2018meta} was proposed. MAL is a framework to create domain-specific languages (DSLs) that defines the generic attack logic codified in languages such as coreLang \cite{coreLang} or powerLang \cite{powerLang}. 

To assess the security state of business or integration processes, we map BPMN to coreLang (a DSL representing general IT concepts) \cite{coreLang}.
Based on this mapping, we automatically transform the concrete process instances into attack models that can be simulated in an attack simulation tool securiCAD \cite{ekstedt2015securi}.
Subsequently, we elaborate more detailed on our applied research method: 

\labeltitle{Methodology}
In this work we design an artefact that provides a security assessment for information systems.
Accordingly, we rely on the principle of Design Science Research (DSR) \cite{Hevner.2004}.
To decide which concrete implementation of DSR we apply, we follow Venable et al. \cite{Venable.2017}, 
a subjectivist, interpretive methodology by developing an artefact that is tailored to the needs of a certain problem.
As we are focusing on just one organization at the moment, we opt for Action Design Research (ADR) \cite{ADR}, which is comprised of four stages, which are illustrated subsequently.

\textbf{Stage 1: Problem Formulation} We face the challenge of large repositories of business and application integration processes between different organizations.
To reduce the risk of unwanted compromise of the connected systems, we want to assess the actual security state within these processes.
Usually, one would scan the systems and perform an automated vulnerability analysis, which is not possible because the processes cover systems of different organizations and vulnerabilities could be introduced in scenario-specific configurations.
Due to the size of the repository, it is also not feasible to follow other approaches that extend BPMN with security related concepts, because it would be too much effort and the process owners might lack the needed security knowledge.
To sum up, we face following challenges in our research:
\begin{itemize}
    \item A large set that hinders (a) a manual security assessment and (b) the reuse of solutions that expect complementing the documented processes.
    \item The processes involve different organizations and scenario-specific configurations.
    Hence, conventional network scans are not an option.
    \item The process owners might have insufficient security background / knowledge.
\end{itemize}

From these challenges, we deduct the following objectives:
\begin{enumerate}[(i)]
    \item \textbf{Processes:} The solution should be able to assess the security of processes notated with BPMN.
    \item \textbf{Automatic:} The assessment should be automatic (\ie to account for a potentially high number of processes).
    \item \textbf{Non-Invasive:} The solution should not demand changes to the existing process documentation.
    \item \textbf{Simulation:} The solution should simulate different applications that are hidden behind interfaces.
\end{enumerate}

\textbf{Stage 2: Building, Intervention, and Evaluation} In stage 2, we follow an IT-dominant approach. On the one hand, we have a team of researchers who develop the artefact presented in \cref{sec:artefact}, while a practitioner continuously provides his feedback on the artefact. The end-users are representatives of the industrial partner developing the processes and others responsible for the security.

\textbf{Stage 3: Reflection and Learning} While in stage 2 a specific solution for a certain problem is developed, stage 3 focuses on the continuous interchange between the researchers and the practitioners. Therefore, we have set up regular meetings in which the recent developments were discussed and mirrored against the industrial partner's reality.

\textbf{Stage 4: Formalization of Learning} The outcomes of stage 2 denote the following contributions of this work:
\begin{itemize}
    \item a BPMN process security classification (focus on process and inter-process, but not conversations),
    \item a mapping for automated translation of BPMN to MAL,
    \item and a prototype for attack simulations.
\end{itemize}

\labeltitle{Outline} We set our approach into context to related work in \cref{sec:relatedWork} and give sufficient background in \cref{sec:background}, before we specify a mapping of BPMN to coreLang in \cref{sec:artefact}.
We evaluate the mapping in \cref{sec:evaluation} and conclude in \cref{sec:conclusion}.

\section{Related work}
\label{sec:relatedWork}
In this section we discuss related work in the intersection of BPMN and security 
in the context of our objectives (i)--(iv), summarized in \cref{tab:related_work} (\enquote{non-invasive} stands for no Meta-model changes required).
Essentially, we found that (a) most of the current approaches require meta-model extensions that add a considerable number of security-related BPMN components, of which many are (b) excessively verbose, and thus difficult to manage (\eg \cite{mulle2011security,salnitri2014modeling,zareen2020security}) or fail to express security concepts in a format that is fully comprehensible to business experts, others propose only a theoretical or descriptive extension (\eg \cite{maines2016adding,meland2012representing}).
%
\begin{table}[tb]
	\centering
	\small
	\caption{\textsc{Bpmn2Mal} in the context of related work}
	\label{tab:related_work}
	\begin{tabular}{p{1.5cm}|cccc}
		\hline
          Literature / objective & \parbox[t]{1.0cm}{Processes (cf. (i))} & \parbox[t]{1.0cm}{Automatic (cf. (ii))} & \parbox[t]{1.7cm}{Non-Invasive (cf. (iii))} & \parbox[t]{1.8cm}{Attack simulation (cf. (iv))} \\
		\hline
		  Threat modeling & \faThumbsUp & \faThumbsODown & \faThumbsODown & \faThumbsODown \\
		  Security modeling & \faThumbsUp & \faThumbsODown & \faThumbsOUp & \faThumbsODown \\
		  Goal modeling & \faThumbsUp & \faThumbsODown & \faThumbsODown & \faThumbsODown \\
		\hline
          \textsc{Bpmn2Mal} & \faThumbsUp & \faThumbsOUp & \faThumbsUp & \faThumbsUp \\
	\end{tabular}

	\centering
	\small
	\faThumbsUp: supported, \faThumbsOUp: partially supported, \faThumbsODown: not supported, \textvisiblespace: n/a

\end{table}
\labeltitle{Threat modeling}
A threat profile security framework was proposed as a BPMN extension by Zareen et al. in \cite{zareen2020security}. The authors leveraged the extension mechanism provided in BPMN 2.0 to model threat-based security requirements and introduced several graphical components for BPMN diagrams.

Meland et al. in \cite{meland2012representing} related the concept of threat modeling to the concept of business process modeling, presenting four different ways for threat specification at designing-time within BPMN. In particular, they discuss the pros and cons of threat representation as error events, escalation events, annotations, and through meta-model extensions.

The found threat modeling approach target objective (i), but do not fulfill objectives (ii--iv)).

\labeltitle{Security modeling} Rodriguez et al. \cite{DBLP:journals/ieicet/RodriguezFP07} propose a non-compliant BPMN meta-model extension including predefined set of high-level cybersecurity requirements (\eg privacy, integrity) into BPMN process diagrams (mainly pool, message flow, data object and activity), enabling business analysts to express their security needs.
The approach was studied for a healthcare process.
Similarly, M\"ulle et al. \cite{mulle2011security} specify a constraint language to formulate security attached via text annotations expressing high-level security requirements.
Consequently, the diagrams become complex and expert-oriented, and thus business experts could find it hard to understand.
In \cite{sang2015bpmn}, Sang et al. also extended the BPMN meta-model with new security elements, which can be also represented within the BPMN diagrams.

A meta-model extension introduced by Brucker et al. \cite{brucker2012securebpmn} with regards to privacy requirements like access control, separation of duties, binding of duty, and need to know.
Beyond the SecureBPMN language, the authors propose a tool to model the security concepts during the modeling phase and also to enforce them at runtime. 
Cherdantseva et al. \cite{cherdantseva2012towards} extended BPMN to include Information Assurance \& Security (IAS) requirements (incl. vulnerabilities, threats, and risks).
Most notably, the work enriches BPMN data-related elements like Data Object as IAS Asset and Message Flow as Vulnerability or Risk.

In \cite{chergui2020towards}, Chergui et al. proposed a BPMN meta-model extension based on the security requirements derived from the cybersecurity ontology in \cite{meland2012representing}.
The extension is fully BPMN compliant and, in contrast to the other works, leverages the BPMN meta-model extension mechanism introduced in BPMN version 2.0.
Also, a web tool was proposed to facilitate collaboration between business and security experts and provided an XML schema extension for integration with the existing BPMN modeler tools.

Maines et al. introduced a comprehensive cybersecurity ontology for specifying security requirements within BPMN, identifying a total of $79$ security concepts \cite{maines2015cyber}. It was later extended \cite{maines2016adding} to represent the BPMN security requirements in a third dimension. However, the extension is described only theoretically.

The security modeling approaches target objective (i) and partially support (iii), but violate objectives (ii) and (iv).

\labeltitle{Security goal modeling} In \cite{salnitri2014modeling}, Salitri et al. proposed a framework to express security requirements in terms of BPMN annotations, called SecBPMN. SecBPMN allows for the description of system information, whereas the security policies are defined through SecBPMN-Q, a query language for BPMN.
The annotated security requirements are derived from the Reference Model of Information Assurance and Security \cite{cherdantseva2013reference} and include accountability, auditability, authenticity, availability, confidentiality, integrity, non-repudiation and privacy.

The goal modeling approach targets objective (i), but does not fulfill objectives (ii--iv)).

\labeltitle{\textsc{Bpmn2Mal}} We focus on a non-invasive modeling of BPMN-based processes without additional security artifacts in BPMN.
Instead \textsc{Bpmn2Mal} strives to automatically map and conduct attack simulations on existing BPMN diagrams and their runtime configurations (\eg remote endpoint interfaces and operations, or scripts and engine configurations). 





\section{Background}
\label{sec:background}
In this section, we give an introduction to MAL, which is the framework in which \texttt{coreLang} is created. \texttt{coreLang} serves as goal for our mapping from BPMN as it is a general representation of IT-dependent systems. 

\subsection{Meta Attack Language}
MAL is a language framework that combines probabilistic attack and defense graphs with object-oriented modeling. It can be used for creating Domain-Specific Languages (DSLs). Such a language defines what information is required and specifies the generic attack logic about the domain studied. We refer to the original paper \cite{johnson2018meta} for a detailed overview of MAL. 

To create a MAL-based language, the first thing is to identify all relevant assets and their associations within a particular domain. Each asset can contain multiple attack steps, representing real threats. One compromised attack step can lead to (represented by "-$>$") a next attack step, where each attack step is of the type OR (represented by "$|$") or AND (represented by "\&"). OR indicates that an attacker can work on this attack step as soon as one of its parent attack steps is compromised, while AND indicates that all its parent attack steps must be compromised for an attacker to reach this step. An asset may also feature defenses (represented by "$\#$"). The sum of attack paths is the attack/defense graph used for the attack simulation. Also, assets can inherit from each other, which means that an inherited asset inherits all attack steps and defenses of its parent asset (unless explicitly stated otherwise).

In Listing \ref{lst:code}, we present a short example of a MAL-based DSL. It contains four modeled assets, the attack step connections, and the connections between assets. In Line 7, the \textit{connect} attack step is of type \textsc{OR}, while in Line 15 is an \textsc{AND} attack step. The \textsc{-$>$} symbol denotes the connected next attack step. Thus, a \textit{phish} on the \texttt{User} leads to \textit{obtain} on the associated \texttt{Password}, and finally to \textit{authenticate} on the associated \texttt{Application}. In Lines 28 to 32, the \texttt{associations} between the assets are defined.

\begin{lstfloat}
\begin{lstlisting}
category System {
  asset Connection {
    | access
      -> app.connect
  }
  asset Application {
    | connect
      -> access
    | authenticate
      -> access
    | guessPwd
      -> guessedPwd
    | guessedPwd [Exp(0.02)]
      -> authenticate
    & access
  }
  asset User {
    | attemptPhishing
      -> phish
    | phish [Exp(0.1)]
      -> pwds.obtain
  }
  asset Password extends Data {
    | obtain
      -> app.authenticate
  }
}
associations {
  Connection[con] * <- Acc -> * [app]Application
  Application[app] 1 <- Cred -> * [pwds]Password
  User[user] 1 <- Cred -> * [pwds]Password
}
\end{lstlisting}
\caption{Example MAL Code\label{lst:code}}
\end{lstfloat}

\subsection{coreLang}
\begin{figure}[bt]
    \centering
    \includegraphics[width=\linewidth]{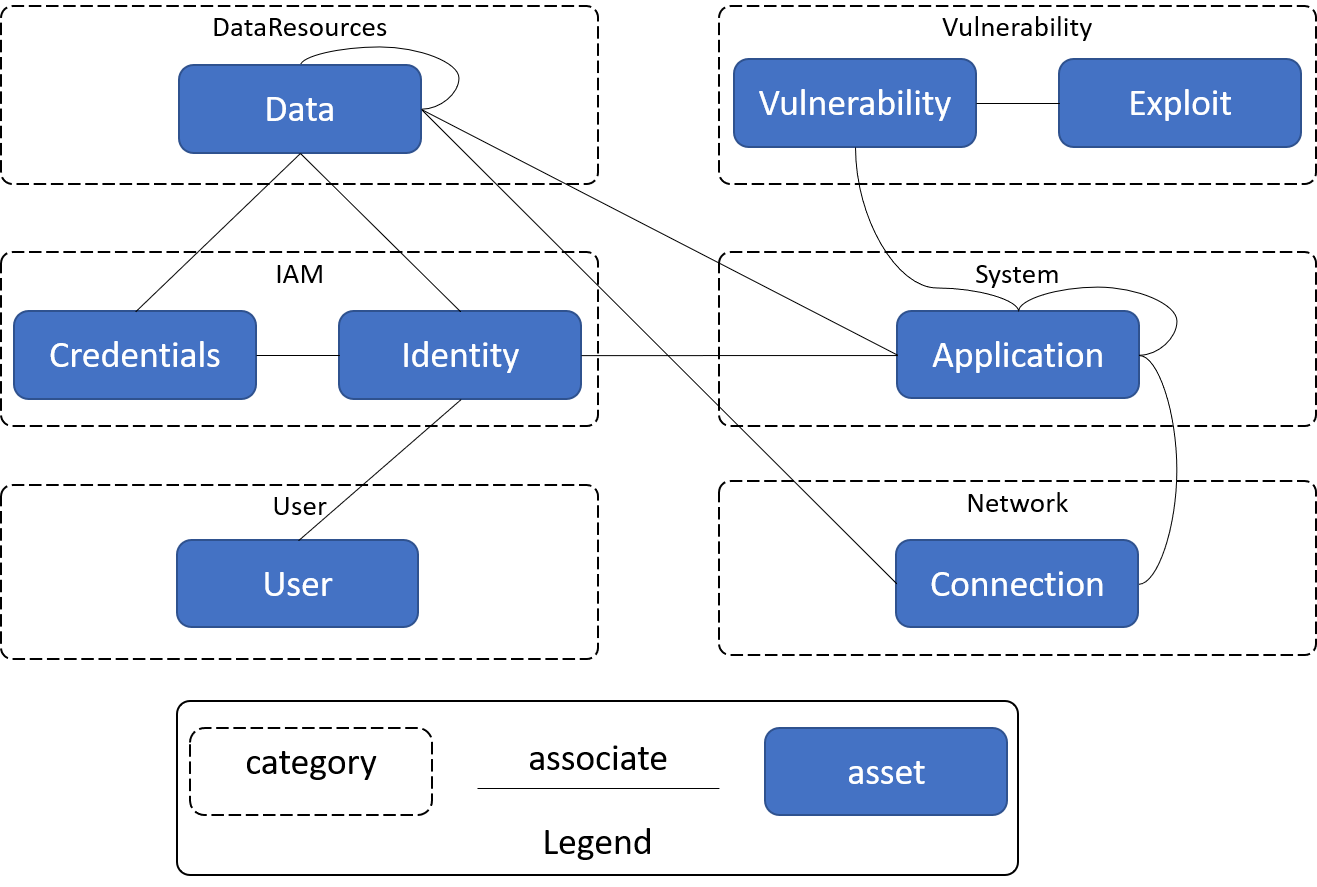}
    \caption{Excerpt of \texttt{coreLang} containing used concepts (adapted from \cite{coreLang}).}
    \label{fig:core}
\end{figure}

As illustrated before, MAL provides the basics to create a threat modeling language from scratch. However, many languages created with MAL share a common set of concepts. To reduce unnecessary redundant work, \texttt{coreLang} \cite{coreLang} was developed. coreLang is comprised of predefined assets that appeared in different languages created with MAL. Thus, this coreLang can serve as starting point to model more domain specific languages or even act as a rudimentary language to model simple environments. \cref{fig:core} presents the overall structure of coreLang. For coreLang, we have identified six different main categories: system, vulnerability, user, IAM (Identity and Access Management), data resources, and networking. Following, we discuss the concepts of coreLang, that we will use later on for our mapping. For more details, we refer to the original publication \cite{coreLang}.

Often, attackers gather initial access to a system by exploiting human factors, represented in our case by the concept of \texttt{User}. These \texttt{User} interact with \texttt{Application} through one or several \texttt{Identity}, which codifies the access rights that the user has. To secure the use of the \texttt{Identity}, \texttt{Credentials} can be applied, which are classically stored as \texttt{Data} and, thus, can be for example encrypted.

Usually, \texttt{Data} is managed by an \texttt{Application}, which represent the main concept used in the following. An \texttt{Application} can run another \texttt{Application}, which can be used to model for example an operation system executing a software. Moreover, \texttt{Application} can communicate with each other via a \texttt{Connection} where they exchange \texttt{Data}.

In an \texttt{Application}, common attack steps are codified. However, if the end-user of \texttt{coreLang} is aware of a certain \texttt{Vulnerability}, which is not included in the common attack steps, they can use the concept of \texttt{Vulnerability} and \texttt{Exploit} to add it to a certain \texttt{Application}.

\section{BPMN attack surface and mapping to coreLang}
\label{sec:artefact}
Hitherto, we have presented MAL and \texttt{coreLang}, which are thought to serve as means for the security analysis. Next, we will elaborate on possible elements of BPMN that can be facilitated as attack surface. Then, the identified surfaces are mapped to \texttt{coreLang}.

\subsection{Attack surface of BPMN}
\label{subsec:surface}
Concepts like vulnerabilities and exploits as part of cyber security attacks captured by MAL and \texttt{coreLang} are crucial to be assessed in BPMN.
While we assume the proper functioning and security checking of the process runtime, we subsequently assess elements of BPMN that can be facilitated as attack surface due to their inherited properties (cf. \cref{tab:surface}).
We consider configuration properties like conditions, expressions, scripts, service calls, and data flow as points of attack. We do not further consider control flow elements as these are managed by the process engine, which we consider as safe, and do not provide an attack surface by themselves.
\begin{table}[tb]
	\centering
	\small
	\caption{Relevant BPMN elements for cybersecurity attacks}
	\label{tab:surface}
\resizebox{\linewidth}{!}{%
\begin{tabular}{p{1.5cm}l|l|l|l|l||l|l}
BPMN     & Sub           & \rotatebox[origin=c]{90}{Condition}          & \rotatebox[origin=c]{90}{Expression}  & \rotatebox[origin=c]{90}{Service Call}  & \rotatebox[origin=c]{90}{Script}   & \rotatebox[origin=c]{90}{Collab. / Proc.} & \rotatebox[origin=c]{90}{Data Flow} \\ \hline
Data object &  &   &      &    &                          & & \faCheck\\
Data Store & &          &      &       &   &  &  \faCheck\\
Message &  &           &      &      &        &  &  \faCheck \\ \hline
Participant / Process &  Pool, Lane &           &      &       &                          & \faCheck &              \\
& Sub-Process &           &      &       &             &                \faCheck &              \\ 
& Event Sub-Process &           &      &       &             &                \faCheck &              \\ \hline
Event & Start                                     &           &      &       &                    &            & \\ 
& Message Start & & &  & & & \faCheck\\ 
& Start Timer                                     &     &  \faCheck    &       &            &           & \\ 
& Start Error                                     &           &      &       &                    &            & \\ 
& Start Compensation                                     &           &      &       &                    &            & \\ 
& Start Parallel                                     &           &      &       &                    &            & \\ 
& Start Escalation                                     &           &      &       &                    &            & \\ 
& Start Condition                                     &    \faCheck       &      &       &                    &            & \\ 
& Start Signal                                     &           &      &       &                    &            & \\ 
& Start Multiple                                     &           &      &       &                    &            & \\ 
\hline
& End                                     &           &      &       &             &                          & \\ 
& Message End    &      \faCheck          &             &     &         &            & \faCheck\\ 
& End Error                                &               &             &               &         &            & \\           
& Cancel End                                &                &             &               &         &            & \\ 
& End Compensation                                &               &             &               &         &            & \\    
& End Signal                                &               &             &               &         &            & \\   
& End Multiple                                &               &             &               &         &            & \\
& Terminate End                                &                &             &               &         &            & \\ 
\hline
Activity & User Task    &               &              &   \faCheck        &  &  & \\  
& Script Task         &           &      &       &        \faCheck     &                                   & \\  
& Service Task   &           &      &   \faCheck    &             && \\  
& Send Task   &           &      &   \faCheck    &             & & \\
& Receive Task   &           &      &  \faCheck     &             &     & \\
& Manual Task  &           &      &    \faCheck   &             &  &\\
& Business Rule Task   &           &  \faCheck    &       &             &                                    & \\
& Call Activity&           &      &     \faCheck  &  &     &  \\
\hline
Gateway & Exclusive & &             &               & & & \\                
& Event-based                                    &                  &             &               &         &            & \\
& Inclusive                        &                         &      &       &                   &            & \\
& Parallel                                    &              &             &               &         &            & \\
& Complex & \faCheck &             &               &         &            & \\ \hline
Sequence Flow & standard / default &           &      &       &                          &  &              \\
 & conditional                                &  \faCheck         &      &       &                          &  &              \\
Association &                                      &           &      &       &                          & &              \\
Message Flow &          &           & \faCheck&       &                          & & \faCheck \\
\end{tabular}%
}
\end{table}

The first property of BPMN we shed light on are conditions. A condition is some kind of threshold that --- if it crossed --- triggers a certain behavior in the process. As thresholds are usually defined by the user, these can be manipulated by the attacker. Alternatively, it is also possible that the attacker changes the input for the condition to (not) trigger it intently. E.g., \textit{Start Condition} is such an event as a process starts if a certain condition is fulfilled, which could have been manipulated by an attacker. Another example is a \textit{Conditional Sequence Flow}, as those include a condition check before continuing, which can be manipulated by an attacker to cause undesired behavior of the process.

Similar to conditions are expressions, which can be more elaborated and are interpreted. Thus, expressions provide a gateway for attackers to manipulate a process to their demands. \eg a \textit{Business Rule Task} is defined by a complex expression that evaluates a set of input parameters and takes a decision based on this input (\eg by a business rule engine).
An attacker can either make changes to the input parameters or --- if they want to create more permanent changes --- manipulate the expressions themselves, and thus exploiting the business rule engine.
Further, there are \textit{Start Timer}, which can contain an expression that regulates when a process is triggered. Here, an attacker could manipulate the starting time to initiate the process more often/seldom as expected.

One main task of Information Systems is to process data. Accordingly, this data is also of interest for attackers, who might want to get access to this data or even to manipulate it. The latter can cause severe issues for the process, as important decisions can be taken based on this data. Therefore, data related elements such as \textit{Data object} or \textit{Message} need special consideration in the security analysis.

Scripts are a powerful tool for developers to easily adopt requested functionality without the need for compiling new versions of programs. This made them very popular among process designers to alter processes to meet the customers' desires. However, a \textit{Script Task} can serve also as attack surface, as the script might be prone to manipulations or the interpreter behind the script has unsolved vulnerabilities.

It is essential for integration processes to provide collaboration mechanisms. As those mechanisms are responsible for organizing the entire process, they need to be in contact with all related elements of the process. Consequently, mechanisms of collaboration are of tremendous interest for an attacker, as the mechanisms can bring the attacker in the position to take over the entire process. Classically, these mechanisms are presented by a \textit{Participant} or a \textit{Pool}. Alternatively, a \textit{(Sub)Process} can also be providing such a means. If an attacker receives access to the application behind the orchestration, they will be in a good position to attack other related activities.

Finally, there are service calls, in which activities are performed outside of the integration process. As the services elaborating on the call can even be outside of the organization or desire human interaction, they are prone to be exploited by an attacker. For example, a \textit{User Task} could be exploited by tricking employees to perform not-safe actions. Considering a \textit{Service Task}, the service can run a outdated server version that will easily allow the attacker to take it over and use it as a starting point to penetrate the process.

For the later, we do not consider those BPMN elements in \cref{tab:surface}, that did not receive a \faCheck, as relevant for our mapping. Either they model a process related behavior that is not of relevance for a static threat model (\eg \textit{Sequence Flow}) or we consider them as already implicitly included in other elements. An example for the latter is the gateway. An attacker could exploit it by manipulating its input parameters. However, the behavior of gateways is managed by the collaboration mechanisms and, thus, represent no additional threat to be considered for our attack simulations.

\subsection{BPMN to coreLang Mapping}
Before, we have determined the relevant elements of BPMN to be mapped to \texttt{coreLang}. Our mapping relies mainly on three mapping rules, that can be summarized as follows:
\begin{enumerate}
    \item Data related BPMN elements will be mapped to \texttt{Data}.
    \item Elements that require user interaction are mapped to \texttt{User}, \texttt{Identity} (the digital representation of the user), and \texttt{Credentials}.
    \item Elements that can be configured or programmed by the process designer, are mapped to \texttt{Application} and \texttt{Connection}, where the latter illustrates the communication between several \texttt{Application}.
\end{enumerate}

However, these rules have exceptions, which we will discuss in the following.

In \cref{tab:mapping_categoryX}, we present our mapping for the \textbf{collaboration} aspects of BPMN to \texttt{coreLang}. Our general interpretation of \textit{Processes} is that they orchestrate the overall happening. Therefore, a tool is needed taking over this task such as a process engine. Based on this observation, we map \textit{Process} and \textit{Sub-Process} to an \texttt{Application}.
%

As MAL-based languages do not have a focus on the process flow but on the static connection between the different assets, there is generally no counterpart for the different \textbf{events} of BPMN in \texttt{coreLang}. However, there are some exceptions: If the event is related to a message (\ie \textit{Message Start} or \textit{Message End}), then there is obviously a communication possible. This communication relates to an exchange of information, which can be manipulated. Therefore, we map \textit{Message Start} and \textit{Message End} to \texttt{Data}. Further, there are \textit{Start Timer} and \textit{Start Condition} that are triggered by an \texttt{Application} that continuously evaluates the underlying conditions.

\begin{table}[tb]
	\centering
	\small
	\caption{Mapping from BPMN to \texttt{coreLang}}
	\label{tab:mapping_categoryY}\label{tab:mapping_categoryX}\label{tab:mapping_data}\label{tab:mapping_connecting}
\resizebox{\linewidth}{!}{%
\begin{tabular}{p{2cm}p{2cm}|l|l|l|l|l|l}
BPMN     & Sub           & \rotatebox[origin=c]{90}{Data}          & \rotatebox[origin=c]{90}{Credentials}  & \rotatebox[origin=c]{90}{Identity}  & \rotatebox[origin=c]{90}{User}      & \rotatebox[origin=c]{90}{Application} & \rotatebox[origin=c]{90}{Connection}\\ 
\hline
Collaboration &  Participant / Process&           &      &       &                          & \faCheck & \faCheck\\
 &  Sub-Process&           &      &       &             &                \faCheck &  \faCheck\\ 
 &  Event Sub-Process       &           &      &       &             &                \faCheck &  \faCheck \\  \hline 
Event 
& Message Start                        &       \faCheck        &              &                  &             &                          & \\ 
& Start Timer      &           &      &       &            &   \faCheck         & \faCheck\\ 
& Start Condition    &           &      &       &  & \faCheck & \faCheck\\ 
& Message End                                 &      \faCheck          &             &               &         &            &  \\ 
\hline
Activity & User Task                        & & \faCheck & \faCheck & \faCheck &                              & \\  
& Script Task  &           &      &       &             &  \faCheck & \faCheck \\  
& Service Task  &           &      &       &             &   \faCheck   & \faCheck\\  
& Send Task   &           &      &       &             &                     \faCheck               & \faCheck \\
& Receive Task   &           &      &       &             &                     \faCheck  & \faCheck\\
& Manual Task                                     &           &      \faCheck & \faCheck     &       \faCheck      &                                    & \\
& Business Rule Task    &           &      &       &             &                     \faCheck               & \faCheck\\
& Call Activity  &           &      &       &             &                     \faCheck   & \faCheck\\ \hline 
Data & Data object &   \faCheck        &      &       &                          & &              \\
 & Data Store&           &      &       &                          & \faCheck &   \faCheck           \\
 & Message &  \faCheck         &      &       &                          &  &              \\ \hline
Connecting  & Sequence Flow (conditional)  &           &      &       &                          & \faCheck &  \faCheck            \\
 &      Message Flow&           &      &       &                          & \faCheck &   \faCheck           \\
\end{tabular}%
}
\end{table}

Next, we discuss the mapping from \textbf{activities} to \texttt{coreLang}. Firstly, BPMN defines \textit{User Task}, where a human is expected to interact with the process. These interactions are a classic attack surface and, thus, we map \textit{User Task} to \texttt{User} in \texttt{coreLang}. Secondly, there are automated activities, such as \textit{Script Task} or \textit{Service Task}. We map both to \texttt{Application}. This is founded in the fact that in a \textit{Script Task} there is an engine (\ie \texttt{Application}) executing a script, while a \textit{Service Task} classically abstracts an interaction with \eg a web-service, which are also executed by an \texttt{Application}.

Regarding the related \texttt{Connection}, we need to differentiate between synchronous and asynchronous calls. Basically, there is one \texttt{Connection} for synchronous calls, because the communication is initiated once. In contrast, there are two \texttt{Connection} for asynchronous calls, as communication gets initiated twice. Accordingly, \textit{Send Task} and \textit{Receive Task} have one \texttt{Connection}, while \textit{Service Task}, \textit{Message Start}, and \textit{Message End} will have one \texttt{Connection}.

BPMN is a process-oriented language. Hence, it provides \textbf{gateways} that describe branching points of the process. In contrast, MAL-based languages codify the static connection between the different assets. Consequently, there is no equivalent in \texttt{coreLang} and just the general connections between the linked activities are transferred.

Often integration processes cope with \textbf{data} by transferring or transforming it. Moreover, to achieve access to data is often the ultimate goal of attackers. In BPMN this concept is modeled by \textit{Data object} and \textit{Message}. As there is no differentiation in \texttt{coreLang}, both are mapped to \texttt{Data}. If a \textit{Data object} is stored outside of the process, then this is modeled by means of a \textit{Data Store}, which can be for example a file system or a database. Obviously, this can be mapped to an \textit{Application} (cf. \cref{tab:mapping_data}).
%
%
For the threat modeling, flows are of no greater importance as indicated before. Nonetheless, there are exceptions in the \textbf{connecting} category of BPMN that require some logic and, thus, are mapped to \texttt{Application} (cf. \cref{tab:mapping_connecting}). In a \textit{Conditional Sequence Flow}, the \texttt{Application} determines based on conditions if a communication takes place, while in a \textit{Message Flow} expressions are evaluated.

%
%

\section{Case Study}
\label{sec:evaluation}

\subsection{\textsc{Bpmn2Mal} Prototype}
\begin{figure}
    \centering
    \includegraphics[width=0.9\linewidth]{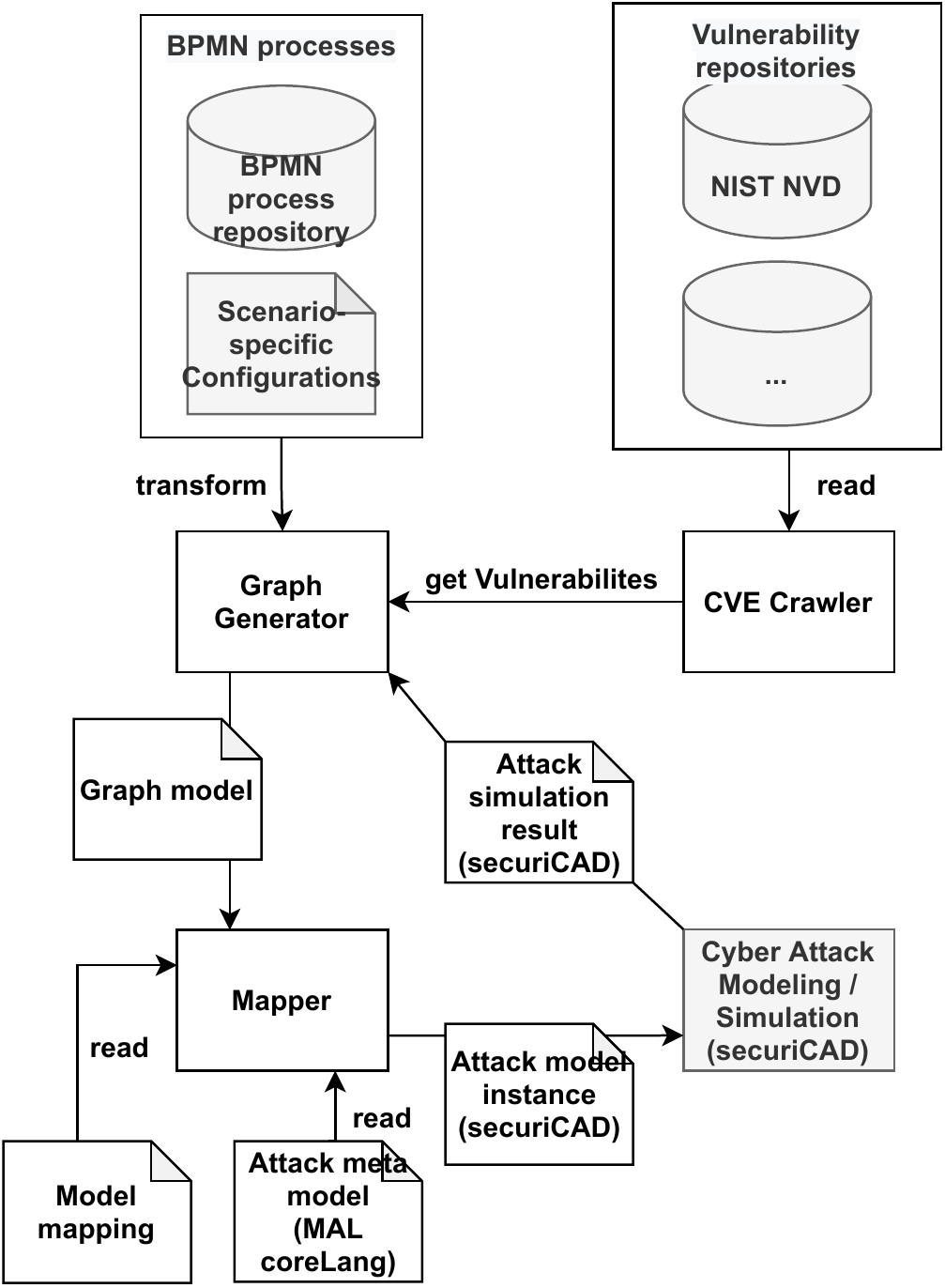}
    \caption{Prototype Architecture}
    \label{fig:toolchain}
\end{figure}

To evaluate our approach, we implemented a prototype\footnote{\textsc{Bpmn2Mal} prototype, visited 6/2021: \url{https://github.com/dritter-hd/bpmn2mal}} (cf. \cref{fig:toolchain}), which has a process model in BPMN as central input. To maintain objective (iii) --non-invasive--, we do not directly elaborate on the process model, but transform the model to a graph representation. This representation includes the overall structure of the process and the attributes that are necessary to perform the mapping from BPMN to \texttt{coreLang}.

To meet objective (iv) --simulation--, which arises from the challenge that we do not know the concrete software specification beyond every interface, we create different graphs that are enriched with respective specifications. Therefore, we provide a list of configurations for each included BPMN element that contains the possibly used libraries and the expected versions. This list is then used to crawl the NIST NVD\footnote{NVD NIST, visited 6/2021: \url{https://nvd.nist.gov/}} for known vulnerabilities of these possible solutions (\eg Apache web server or nginx) and compute the time-to-compromise following McQueen et al. \cite{McQueen}. This information is added to the graph representation and additionally, the different solutions are combined. In other words, we prepare different graphs that to some degree contain \eg Apache web server, while other graphs contain nginx or a combination of both.

These different graphs are then transformed into securiCAD models \cite{ekstedt2015securi} that are instances of \texttt{coreLang} \cite{coreLang} following the mapping described in \cref{sec:artefact}. This is extending previous work \cite{edoc}, which focused on automatically creating MAL languages, while here we are creating concrete instances of such languages. For each of the graphs, several simulations are performed giving the simulated attacker different attack surfaces (cf. \cref{subsec:surface}). Then, the results of the different simulations get aggregated and are returned back to the graph representation to finally visualize the critical activities in the process model.

%
%

\begin{figure*}
    \centering
    \includegraphics[width=\textwidth]{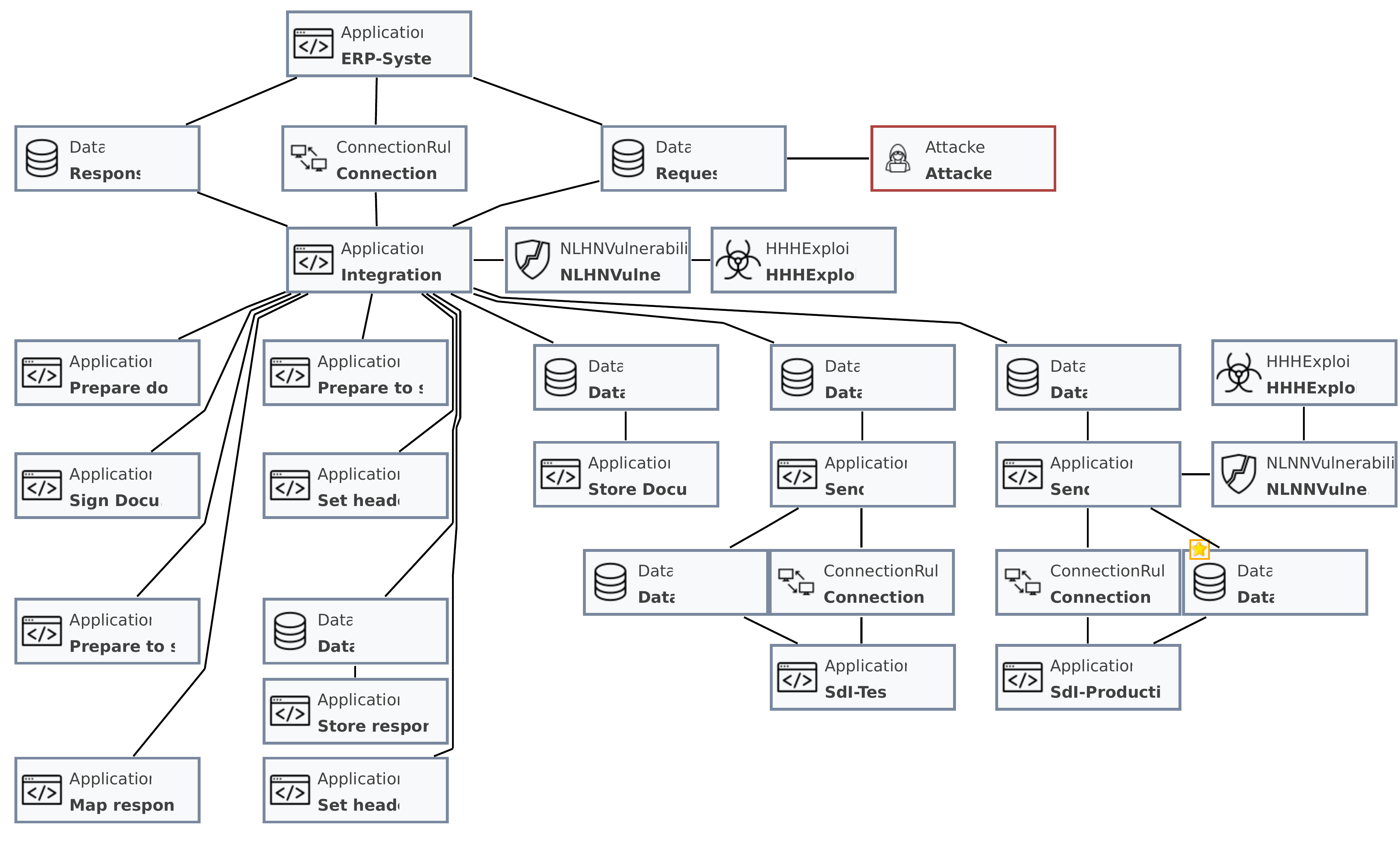}
    \caption{Threat model representation of Fig. \ref{fig:cdd}}
    \label{fig:threatmodel}
\end{figure*}
\subsection{Motivating example (revisited): Italy invoicing}
With \cref{fig:cdd}, we introduced an example invoicing integration process. Facilitating the mapping presented in \cref{sec:artefact}, we translate this integration process to the threat model presented in \cref{fig:threatmodel}. To not clutter \cref{fig:cdd}, it does not contain explicit data objects and related associations, which usually would be included. Accordingly, the respective \texttt{coreLang} representation does not contain explicit \texttt{Data} \includegraphics[scale=0.15]{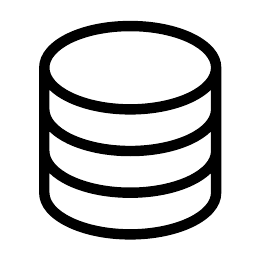} either.

For greater clarity, \cref{fig:threatmodel} contains only a subset of the possible vulnerabilities \includegraphics[scale=0.15]{img/Vulnerability} and exploits \includegraphics[scale=0.023]{img/Exploit}. More concretely, just those that are needed to enable the attacker to write the data transmitted to \textit{SdI-Production}, which is symbolized by the star on \texttt{Data} \includegraphics[scale=0.15]{img/Data}. The top of the threat model describes the communication between the \textit{ERP-System} and the \textit{Integration System}. In the lower left part, the different \textit{Script Task} and \textit{Service Task} take place, which mainly process the document. The lower right part describes the communication with the SdI for the test and the productive system.

Next, we discuss a penetration, where an attacker can perform a Man-In-The-Middle (MITM) attack on the request (\ie \texttt{Data} \includegraphics[scale=0.15]{img/Data}) sent from \textit{ERP-System} (\ie \texttt{Application} \includegraphics[scale=0.15]{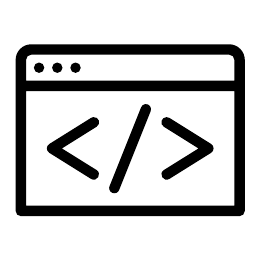}) to \textit{Integration System} (\ie \texttt{Application} \includegraphics[scale=0.15]{img/Application}) via a \texttt{Connection} \includegraphics[scale=0.15]{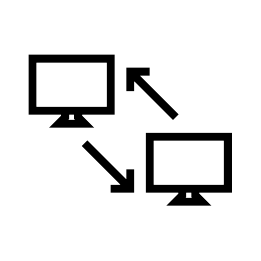}, \eg because an older version of encryption is used. This enables the attacker to inject code into a component used by the process engine (CVE-2021-23337)\footnote{Lodash vulnerability used in Camunda, visited 6/2021: \url{https://nvd.nist.gov/vuln/detail/CVE-2021-23337}}. As this vulnerability allows to execute arbitrary code, the attacker can alter the data sent to the \textit{Service Task} communicating with SdI. If this \textit{Service Task} is run by a miss-configured Apache Tomcat (CVE-2020-1938)\footnote{Apache Tomcat vulnerability, visited 6/2021: \url{https://nvd.nist.gov/vuln/detail/CVE-2021-25329}}, the attacker is again able to execute their desired code. Hence, it is possible to \texttt{write} on the related data sent to SdI and the attacker reached their goal.

\begin{figure*}
    \centering
    \includegraphics[width=\textwidth]{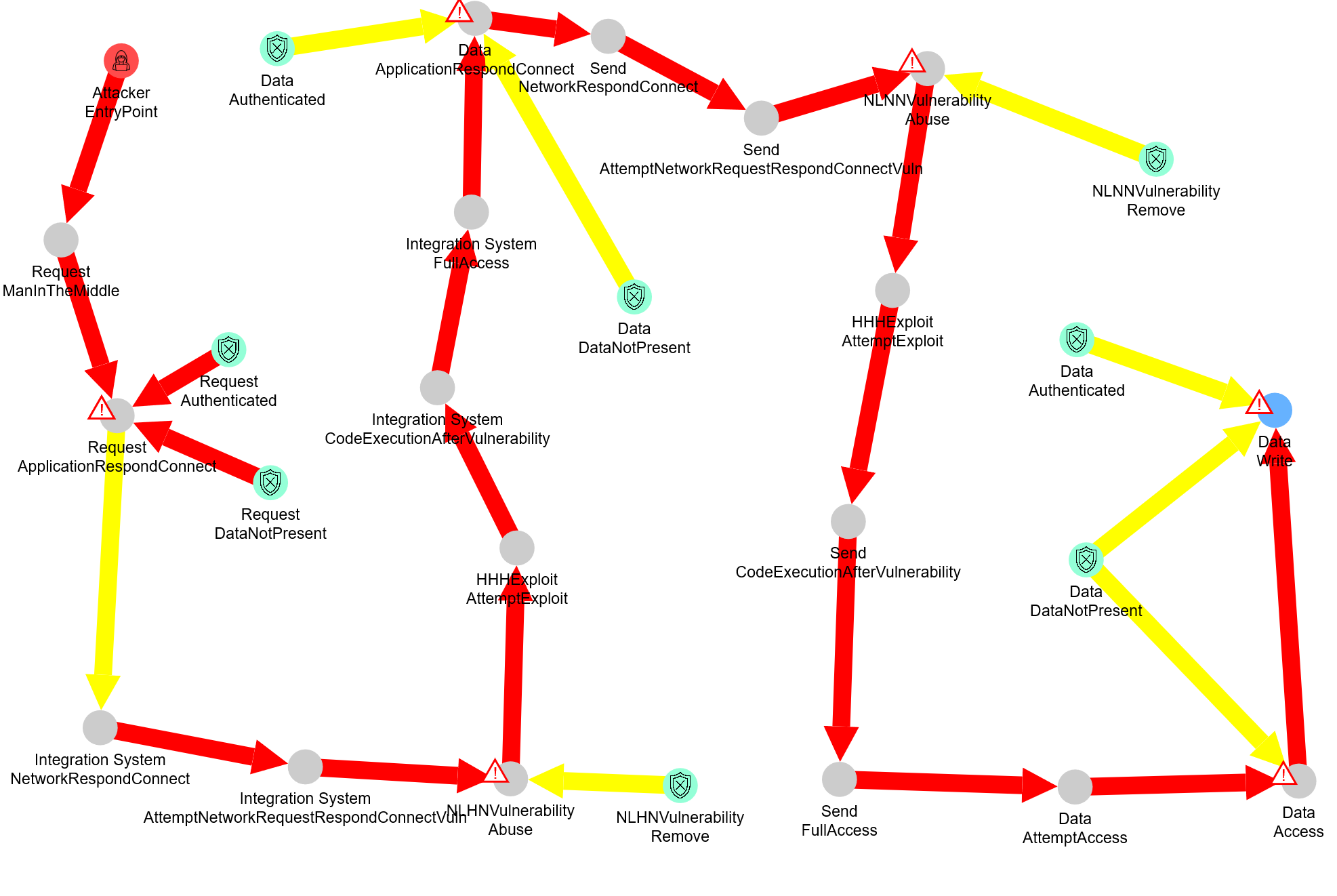}
    \caption{Visualization of the attack path to reach \texttt{Data.Write} sent to \texttt{SdI-Production}}
    \label{fig:attackpath}
\end{figure*}

Now, we are able to perform the attack simulation in securiCAD \cite{ekstedt2015securi} to analyze what path the attacker takes through the process and determine possible countermeasures (cf. \cref{fig:attackpath}).
The analysis shows that the actual design of the process can be considered as secure, as the described attacker achieves a success rate of 8\% after 100 days of penetrating the system.

Nonetheless, some room for improvement was identified. To stop the attacker early, one can require that data transmissions are just accepted if the sender is properly authenticated, thus MITM would not be possible anymore. Another option, which would stop the attacker later, would be to patch the vulnerable applications. Then, these would not be exploitable anymore (at least for exploits related to these specific vulnerabilities).

\subsection{Discussion}
To reflect on our mapping decisions, we discussed them with an experienced threat modeler, who is also familiar with the concepts of \texttt{coreLang}. We explained the integration process in \cref{fig:cdd} and showcased the resulting threat model in \cref{fig:threatmodel}. Overall, the expert agreed with our decisions. However, he raised some points that we will discuss further.

In our mapping, we opted to map every \textit{Script Task} to a single \texttt{Application}, because we assumed that every script could contain a vulnerability that might not be present in other scripts. Alternatively, one could combine \textit{Script Task} by the underlying technology, \eg javascript, groovy, python, etc. Then, the focus would be more on the assumption that the vulnerability can be found in those technologies. Lastly, one could argue that there is no need for an explicit modeling of scripts since those are executed by the process engine. Discussing these three possibilities, we concluded that the first solution is the better one, as it contains information which scripts are corrupted, the data exchange between them is more obvious, and it is closer to the original process model.

Another point brought up was to split up the \textit{Process / Participant} into several \texttt{Applications} that represent the different functionalities. In our case, this would result in splitting the \textit{Integration System} into two endpoints to receive the request and to send the response, and a component managing the process execution. However, we neglect this additional differentiation, as we do not see any added value and more elements would be presented in the final model.

Taking a closer look at the attack simulations presented before, we recognize that for a successful penetration of the process, the \textit{Integration System} is crucial. To reach the other parts of the process, the easiest and shortest way usually facilitates this central component. This observation can be generalized for all integration processes that make use of a process engine. Thus, we can conclude that the security engineers should put emphasis on that component.

Another observation is related to the positioning of the attacker. In our example above, the attacker can perform a MITM on the request. However, there are further attacks possible, like eavesdrop, that will result in different outcomes of what is accessible to the attacker and what is not. Moreover, the attacker could be (theoretically) placed on each and every element of the process. However, this is caused by the design of MAL itself. To overcome this, MAL could include an indicator, which assets and which attacks are the most likely attack surfaces.

This complexity is further increased by performing simulations for different configurations of the assets (\ie a \textit{Service Task} can either be executed by an Apache or an nginx http server). Each possible combination increases the number of needed simulations exponentially. Consequently, a solution is needed that reduces this number. One approach could be to restrict the possible configurations to realistic ones. For example, one can assume that within one organization (\ie \textit{Participant / Pool}), it is very likely that only one kind of a certain application for one purpose is deployed and only the versions might differ on a certain interval. Another possible assumption is to rely on usage statics of certain applications (\eg http server) and to solely consider those that are used to a certain degree.

One aspect that can be improved in future is the fact that all our integration processes at hand only contain a subset of the possible BPMN elements. Especially, the integration processes are completely automatized and, hence, there is no interaction with humans expected. But the users of information systems are often used as attack surface to gain access to those systems. Moreover, human related security properties can differentiate significantly among organizations --- and even in sub-entities of organizations. This cannot be codified in a static language such as \texttt{coreLang}. Therefore, an approach is needed that assess the security behavior in organizations (\eg \cite{SBA}) and includes this information into the performed attack simulations.

\section{Conclusion}
\label{sec:conclusion}
To conclude our work, we revisit our objectives (i)--(iv). Objective (i) focused on the ability to assess the security of processes represented in BPMN.
To achieve this objective, we created a mapping from BPMN to \texttt{coreLang}, which enables us to analyze the security of the respective process. 

Objective (ii) suggested an automatic security assessment.
We partly satisfy this objective in the current proposal. In our solution, the BPMN model is translated automatically to the format needed, however the simulations need to be manually started as securiCAD is missing a command interface at the moment. This would allow us to automatize this part as well.
An alternative could be to use robotic process automation tools to 
automate the user interaction.

We have accomplished objectives (iii) and (iv). For objective (iii), we simply take the process model and enrich a graph representation of it with the needed security information, thus, no changes to the original model are necessary.
For objective (iv), we use the information from the CVE databases to simulate different applications where we are not able to gain deeper knowledge of their characteristics.

A potential next step would be to conduct a user study by having the attack simulations results reviewed and categorized.
It would be interesting to see how easy the findings are to fix. We assume that some findings would be; 1) a quick fix that can just be implemented without any further analysis, 2) doable but with some constraints and perhaps a more in-depth assessment, and 3) nearly impossible to do anything about without large efforts.

Further, we believe that an interesting study would be to analyze a set of old process implementations, review what security-related changes these processes have gone through and re-analyze the same processes in their new state. This, in order to see if our suggested approach did pick up issues that were fixed and also examine why other issues haven't been mitigated yet. 

\labeltitle{Acknowledgements} We thank Rafael L\"udke, Erik Henriksson and Klas Engberg for implementation support, and Joar Jacobsson for valuable discussions on the BPMN to MAL mapping.
This project has received funding from the Swedish Energy Agency.

\bibliographystyle{abbrv}
\bibliography{mybib}








\end{document}